\documentclass[3p,times,procedia,sort&compress]{elsarticle}
\usepackage{nupha_ecrc}

\volume{00}

\firstpage{1}

\journalname{Nuclear Physics A}

\runauth{M. McNelis, D. Bazow and U. Heinz}

\jid{nupha}

\jnltitlelogo{Nuclear Physics A}

\usepackage{amssymb}
\usepackage{lineno}
\usepackage{graphicx}
\usepackage{bm}
\usepackage{amsmath,latexsym}
\usepackage[usenames]{color}
\usepackage{subfigure}
\usepackage{subfigure}
\usepackage{slashed}
\usepackage{multirow,array}
\usepackage{mathtools}
\usepackage{mathrsfs}
\usepackage[colorlinks=false,linktocpage=true]{hyperref}
\usepackage{hyperref}
\usepackage[utf8]{inputenc}
\usepackage{lipsum}
\usepackage[rightcaption]{sidecap}

\usepackage{color}
\definecolor{darkblue}{RGB}{0,0,196}
\definecolor{darkred}{RGB}{196,0,0}

\biboptions{comma,sort&compress}
\usepackage[figuresright]{rotating}

\newcommand{\be}{\begin{equation}}
\newcommand{\ee}{\end{equation}}
\newcommand{\bea}{\begin{eqnarray}}
\newcommand{\eea}{\end{eqnarray}}
\newcommand{\beal}{\begin{align}}
\newcommand{\enal}{\end{align}}
\newcommand{\bs}{\begin{subequations}}
\newcommand{\es}{\end{subequations}}
\newcommand{\besp}{\begin{split}}
\newcommand{\eesp}{\end{split}}
\newcommand{\x}{\hat{x}^\mu}

\newcommand{\ene}{\mathcal{E}}

\newcommand{\Pl}{\mathcal{P}_L}

\newcommand{\Peq}{\mathcal{P}_{\text{eq}}}

\newcommand{\PL}{\mathcal{P}_L}
\newcommand{\Pperp}{\mathcal{P}_\perp}
\newcommand{\Wperp}{W^\mu_{\perp z}}
\newcommand{\piperp}{\pi^\munu_{\perp}}
\newcommand{\munu}{{\mu\nu}}
\newcommand{\eq}{\mathrm{eq}}

\begin{document}

\begin{frontmatter}

\dochead{XXVIIth International Conference on Ultrarelativistic Nucleus-Nucleus Collisions\\ (Quark Matter 2018)}

\title{Viscous hydrodynamics for nonconformal anisotropic fluids\tnoteref{tn0}}
\tnotetext[tn1]{Supported by DOE (award no.\,DE-SC0004286 and BEST Collaboration) and NSF (JETSCAPE Collaboration, ACI-1550223).}

\author[osu]{Michael McNelis\footnote{Presenter. E-mail: mcnelis.9@osu.edu.}}
\author[osu]{Dennis Bazow}
\author[osu,cern,emmi]{Ulrich Heinz}
\address[osu]{Department of Physics, The Ohio State University, Columbus, OH 43210, USA}
\address[cern]{Theoretical Physics Department, CERN, CH-1211 Gen\`eve 23, Switzerland}
\address[emmi]{ExtreMe Matter Institute (EMMI), GSI Helmholtzzentrum f\"ur Schwerionenforschung, 
                          Planckstrasse 1, D-64291 Darmstadt, Germany\\[-5ex]}

\begin{abstract}
A new formulation of (3+1)-dimensional anisotropic hydrodynamics is presented that accounts nonperturbatively for the large longitudinal-transverse pressure anisotropy and bulk viscous pressure in heavy-ion collisions. The initialization of the anisotropic hydrodynamic stage is discussed, and a comparison to standard viscous hydrodynamics for (0+1)-dimensional Bjorken expansion is presented.
\end{abstract}


\end{frontmatter}

\section{Introduction}
\label{sec:introduction}

Relativistic viscous hydrodynamics has been successful in describing the evolution of the quark-gluon plasma phase in heavy-ion collisions. It is the centerpiece of state-of-the-art hybrid models, which describe a wide variety of experimental data for different collision systems. However, $2^{\text{nd}}$ order viscous hydrodynamics relies on the assumption that the shear stress $\pi^\munu$ and bulk viscous pressure $\Pi$ are small compared to the thermal pressure. This is questionable at early times and close to hadronization: strongly anisotropic expansion in heavy-ion collisions drives large longitudinal-transverse pressure anisotropies $\PL{-}\Pperp$ at early times, and the quark-hadron phase transition can generate a large bulk viscus pressure $\Pi=(2\Pperp{+}\PL)/3-\Peq$ towards the end of the fluid dynamic stage. In this contribution we present a new formulation of anisotropic hydrodynamics that evolves the longitudinal and transverse pressures $\PL$ and $\Pperp$ at leading-order. Only the smaller residual shear stresses are treated perturbatively. For systems without conserved charges a full account of the formalism is presented in \cite{McNelis:2018jho}; here we only summarize its essential features but add a discussion of how to initialize the anisotropic parameters that are necessary for computing the initial values of the anisotropic transport coefficients. A brief comparison of the evolution of the pressure anisotropy $\PL/\Pperp$ in (0+1)-dimensional Bjorken expansion between our new approach and standard $2^{\text{nd}}$ order viscous hydrodynamics concludes these proceedings. 

\section{Anisotropic hydrodynamic equations}
\label{sec2}

For anisotropic hydrodynamics the energy-momentum tensor is decomposed as \cite{Molnar:2016vvu}
\be
\label{eq1}
  T^\munu = \ene\,u^\mu u^\nu + \PL \, z^\mu z^\nu - \Pperp\,\Xi^\munu  + 2\,W^{(\mu}_{\perp z} z^{\nu)} + \piperp,
\ee
where $u^\mu$ is the fluid velocity, $z^\mu$ reduces to the longitudinal unit vector in the local rest frame (LRF), $\Xi^\munu = g^\munu - u^\mu u^\nu + z^\mu z^\nu$ projects on the transverse directions in the LRF, and $\ene$, $\PL$, and $\Pperp$ are the LRF energy density, longitudinal and transverse pressure, respectively. The residual shear stress components, given by the longitudinal-momentum diffusion current $\Wperp$ and the transverse shear stress tensor $\piperp$, are assumed to be smaller than $\PL$ and $\Pperp$.

Ten equations are needed to evolve $T^\munu$. The energy-momentum conversation laws evolve $\ene$ and $u^\mu$, with the equilibrium pressure $\Peq(\ene)$ given by the QCD equation of state (EoS). Six relaxation equations for the dissipative terms $\PL{-}\Peq$, $\Pperp{-}\Peq$, $\Wperp$, and $\piperp$ describe the competition between global expansion (driving the system away from local equilibrium and momentum isotropy) and interactions among the microscopic constituents (driving the system closer to local equilibrium). Hydrodynamics being an effective field theory for the long-distance dynamics of multiparticle systems \cite{Baier:2007ix}, the structure of these equations is universal, with specific medium properties encoded in the EoS and a set of transport coefficients. In \cite{McNelis:2018jho} the Boltzmann equation in the relaxation-time approximation for a system of weakly interacting quasiparticles with a medium dependent mass $m(T)$ is used to derive these relaxation equations, including one for a mean field $B$ needed for thermodynamic consistency \cite{Gorenstein:1995vm}. We expand the distribution function as $f = f_a + \delta\tilde f$ where $f_a(x,p) = \exp\Bigl[-\sqrt{m^2 + p_{\perp,\mathrm{LRF}}^2 / \alpha_\perp^2(x) + p_{z,\mathrm{LRF}}^2 /\alpha_L^2(x)} \Big/ \Lambda(x) \Bigr]$ is the leading-order anisotropic distribution and $\delta\tilde f$ (for which we use a 14-moment approximation) generates the residual dissipative corrections. The momentum anisotropy parameters $\alpha_{L,\perp}(x)$ are matched to $\mathcal{P}_{L,\perp}(x)$ dynamically via generalized Landau matching conditions \cite{Tinti:2016bav, Molnar:2016gwq, McNelis:2018jho}. Similarly, the effective temperature $\Lambda(x)$ is matched dynamically to the energy density $\ene(x)$ which defines the local temperature $T(x)$ through the EoS. This yields a relation $T(\Lambda, \alpha_\perp, \alpha_L)$. 

\begin{figure}[t!]
\centerline{
\includegraphics[width=0.8\linewidth]{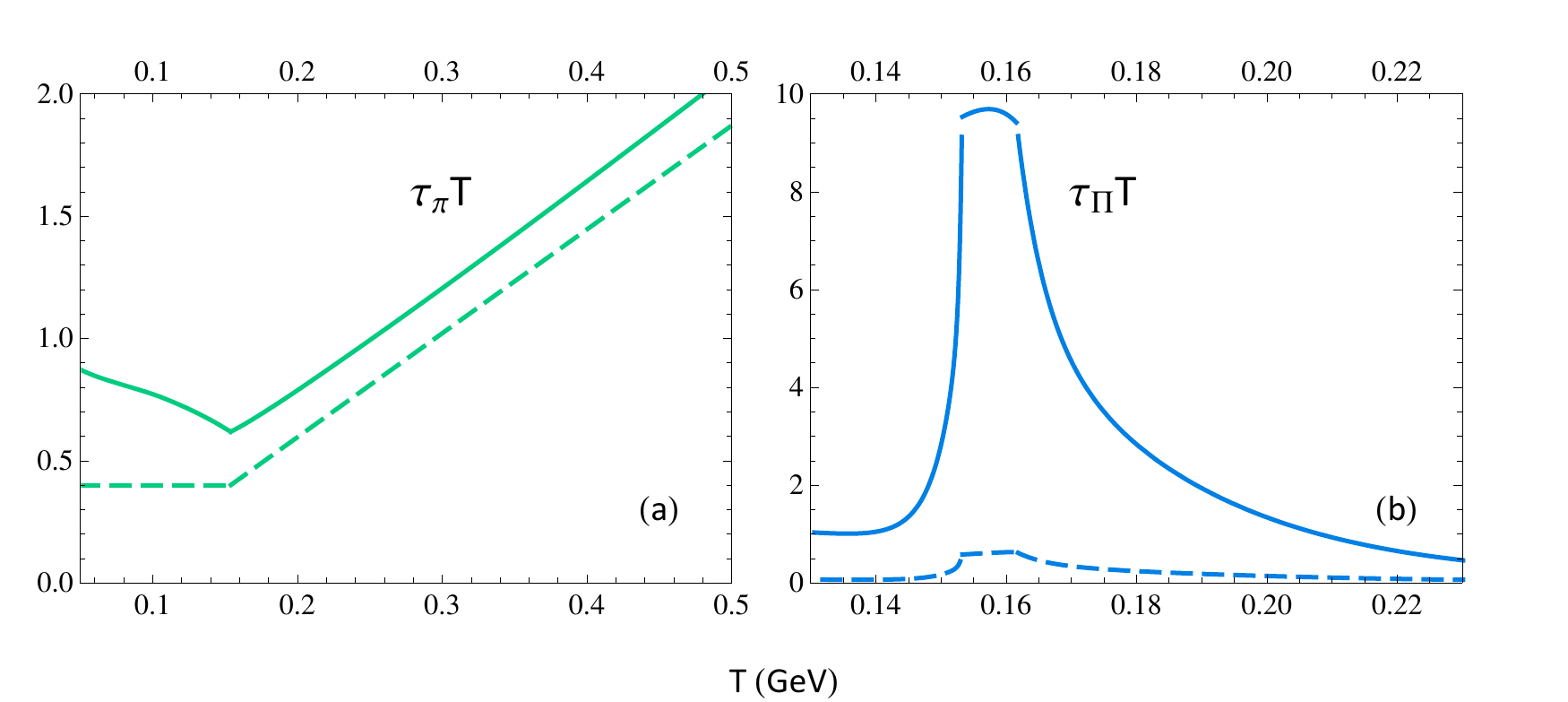}\\[-2.5ex]
}
\caption{
The dimensionless shear and bulk relaxation times computed in the quasiparticle model (solid) and for a nearly massless Boltzmann gas (dashed) as functions of temperature. The temperature dependences of $\eta/s$ and $\zeta/s$ are taken from the Bayesian analysis presented in \cite{Bernhard:2016tnd} while the functions $\beta_{\pi,\Pi}(T)$ for the two models are given in Eqs.~(82) and (E17) of Ref.~\cite{McNelis:2018jho}.
}
\label{fig1}
\vspace*{-3mm}
\end{figure}

After using the generalized Landau matching conditions to eliminate all derivatives of the microscopic anisotropy parameters $\Lambda, \alpha_\perp, \alpha_L$ and the mean field $B$ in terms of the macroscopic hydrodynamic variables $\ene$, $\Pperp$ and $\PL$, a set of purely macroscopic evolution equations is found, with source terms that describe the generation of dissipative flows in terms of hydrodynamic gradients multiplied by transport coefficients \cite{McNelis:2018jho}. Ref.~\cite{McNelis:2018jho} calculates the transport coefficients within the quasiparticle kinetic model as functions of the microscopic parameters $\Lambda, \alpha_\perp, \alpha_L$ which can be obtained (in this model) from $\ene$, $\Pperp$ and $\PL$; eventually, they should be computed directly from QCD.

The relaxation equations contain two relaxation times, $\tau_\pi$ and $\tau_\Pi$. In kinetic theory, they are proportional to the shear and bulk viscosities: $\tau_\pi = (\eta / s) \times (s/\beta_\pi)$ and $\tau_\Pi =  (\zeta / s) \times (s/\beta_\Pi)$. The specific viscosities $(\eta/s)(T)$ and $(\zeta/s)(T)$ are modeled phenomenologically \cite{Bernhard:2016tnd}. In kinetic theory the entropy density $s(T)$ and the coefficients $\beta_{\pi,\Pi}(T)$ are given by thermodynamic integrals over the distribution function \cite{Denicol:2012cn}. Commonly used expressions for $\beta_{\pi,\Pi}$ in the limit $m/T \ll 1$ \cite{Denicol:2014vaa} are found to deviate strongly from their quasiparticle model counterparts where typically $m/T \gtrsim 1$ if the model is tuned to reproduce the lattice QCD EoS \cite{Tinti:2016bav}. Fig.~\ref{fig1} shows the dimensionless relaxation times $T \tau_{\pi,\Pi}$ in the quasiparticle model (solid) and small mass limit (dashed). While the shear relaxation times are similar in the two cases, leading to similar transient dynamics for the shear stress during the early collision stages, the bulk relaxation times differ by an order of magnitude around $T_c\simeq 154$\,MeV. While both models show critical slowing down of the bulk viscous dynamics, this effect is dramatically enhanced in the quasiparticle model tuned to the lattice QCD EoS. The corresponding delay of the medium's response to the scalar expansion rate limits the overall magnitude of the bulk viscous pressure $\Pi$ \cite{McNelis:2018jho}. This could have important phenomenological consequences for Bayesian statistical constraints on $(\zeta/s)(T)$, a full investigation of which may require promoting $T \tau_\Pi$ in Bayesian fits to a phenomenological function of temperature. 

\section{Initialization of anisotropic hydrodynamics}
\label{sec3}

For the calculation of transport coefficients from the quasiparticle model, the microscopic parameters $(\Lambda, \alpha_L, \alpha_\perp)$ must be calculated from the macroscopic entities $(\ene,\Pl,\Pperp)$. This also requires knowledge of the mean field $B$ \cite{McNelis:2018jho} for which we solve an evolution equation in parallel to the relaxation equations for the dissipative flows \cite{McNelis:2018jho}. Like the latter, $B$ requires an initial condition. However, given an EoS $\Peq(\ene)$, not every initial condition for $(\ene,\PL,\Pperp,B)$ can be successfully described within the quasiparticle kinetic theory because the latter features positive definite values for the kinetic pressure contributions. A discussion of possible initializations of $B$ is therefore of interest.   

\begin{figure}[t!]
\centerline{
\includegraphics[width=0.9\linewidth]{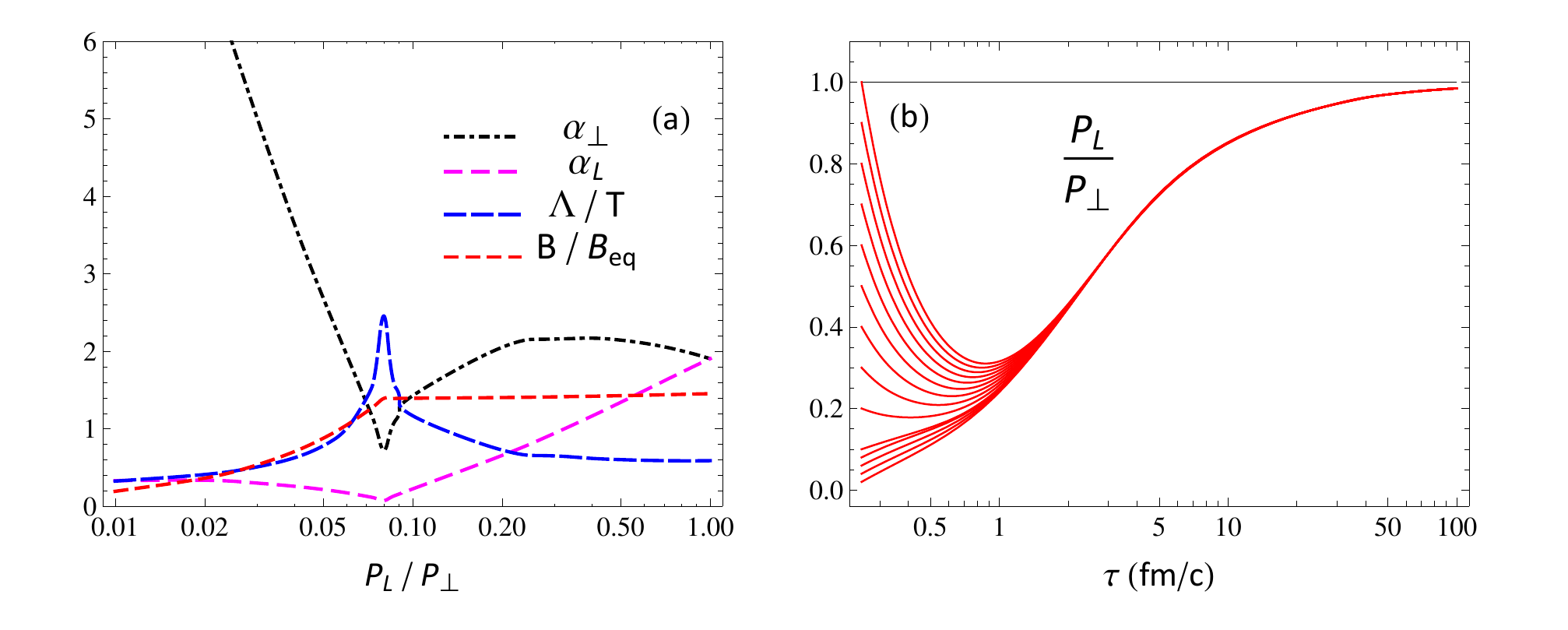}\\[-1.5ex]
}
\caption{
Initial values of the parameters $(\Lambda, \alpha_L, \alpha_\perp)$ and the mean field $B$ as functions of the initial pressure ratio $\PL /\Pperp$ for fixed bulk viscous pressure $\Pi$ (a) and the subsequent Bjorken evolution of $\PL/\Pperp$ (b), for an initial temperature $T_0=0.5$\,GeV at $\tau_0 = 0.25$\,fm/$c$.
}
\label{fig2}
\vspace*{-3mm}
\end{figure}
%

Let us assume (0+1)-dimensional Bjorken expansion and initial values $T^\munu_0=T^\munu(\tau_0)$ for the energy-momentum tensor provided by some conformal pre-hydrodynamic evolution model involving massless degrees of freedom. Matching this to the anisotropic hydrodynamic form (\ref{eq1}) with a QCD EoS, the starting values for $\ene,\PL,\Pperp$ can be obtained from $T^\munu_0$. Since the QCD EoS breaks conformal invariance this matching results in a non-zero initial bulk viscous pressure $\Pi_0$ which is independent of the pressure ratio $\PL/\Pperp$. 
Writing the initial mean field as $B = B_\eq + \delta B$, the equilibrium part $B_\eq(T_0)$ is determined by thermodynamic consistency from the initial temperature $T_0$, provided by the EoS $\ene(T)$. If one assumes that $\delta B$ evolves on a much longer time scale than $\tau_\Pi$ one can set $\delta B \approx (3 \tau_\Pi \dot{m} \Pi) / (m{-}4\tau_\Pi \dot m)$ \cite{McNelis:2018jho}. The microscopic parameters $X = (\Lambda, \alpha_L, \alpha_\perp)$
are then obtained by solving $\Bigl(\ene^{(k)}(X){-}\ene{+}B, \, \PL^{(k)}(X){-}\PL{-}B, \, \Pperp^{(k)}(X){-}\Pperp{-}B\Bigr)=0$ where the superscript $(k)$ denotes the kinetic contributions \cite{McNelis:2018jho}. The results are shown in Fig.~\ref{fig2}a. For $\PL/\Pperp \lesssim 1$, the mean field contributes little to this equation ($B/\Peq \sim -\, 0.1$ at high temperatures), and a simple trend emerges: $\Lambda \approx$ const., $\alpha_L$ decreases, $\alpha_\perp$ increases slightly, and $B \approx$ const. As $\PL / \Pperp$ approaches the value 0.08, $\alpha_L$ approaches zero (leading to zero longitudinal kinetic pressure), and $\alpha_\perp$ also starts to decrease. For $\PL / \Pperp < 0.08$, the root-finding algorithm breaks down because $\PL^{(k)} = \PL + B$ starts to become negative. To keep $\PL^{(k)}$ in bounds one must give up the adiabatic approximation for $\delta B$ introduced above. The solution shown in Fig.~\ref{fig2} assumes that for $\PL / \Pperp{\,\rightarrow\,}0$ the initial mean field $B$ approaches zero linearly.

Figure~\ref{fig2}b shows the ensuing time evolution of $\PL / \Pperp$. The early-time dynamics is insensitive to the detailed assumptions made in the initialization of the mean field $B$ since $\delta B$ is only a small viscous correction to the macroscopic pressures. One notes that after about 1\,fm/$c$ the system has lost its memory of the initial conditions and the pressure anisotropy converges onto a hydrodynamic attractor.

\begin{SCfigure}[1][h]
\includegraphics[width=0.42\linewidth]{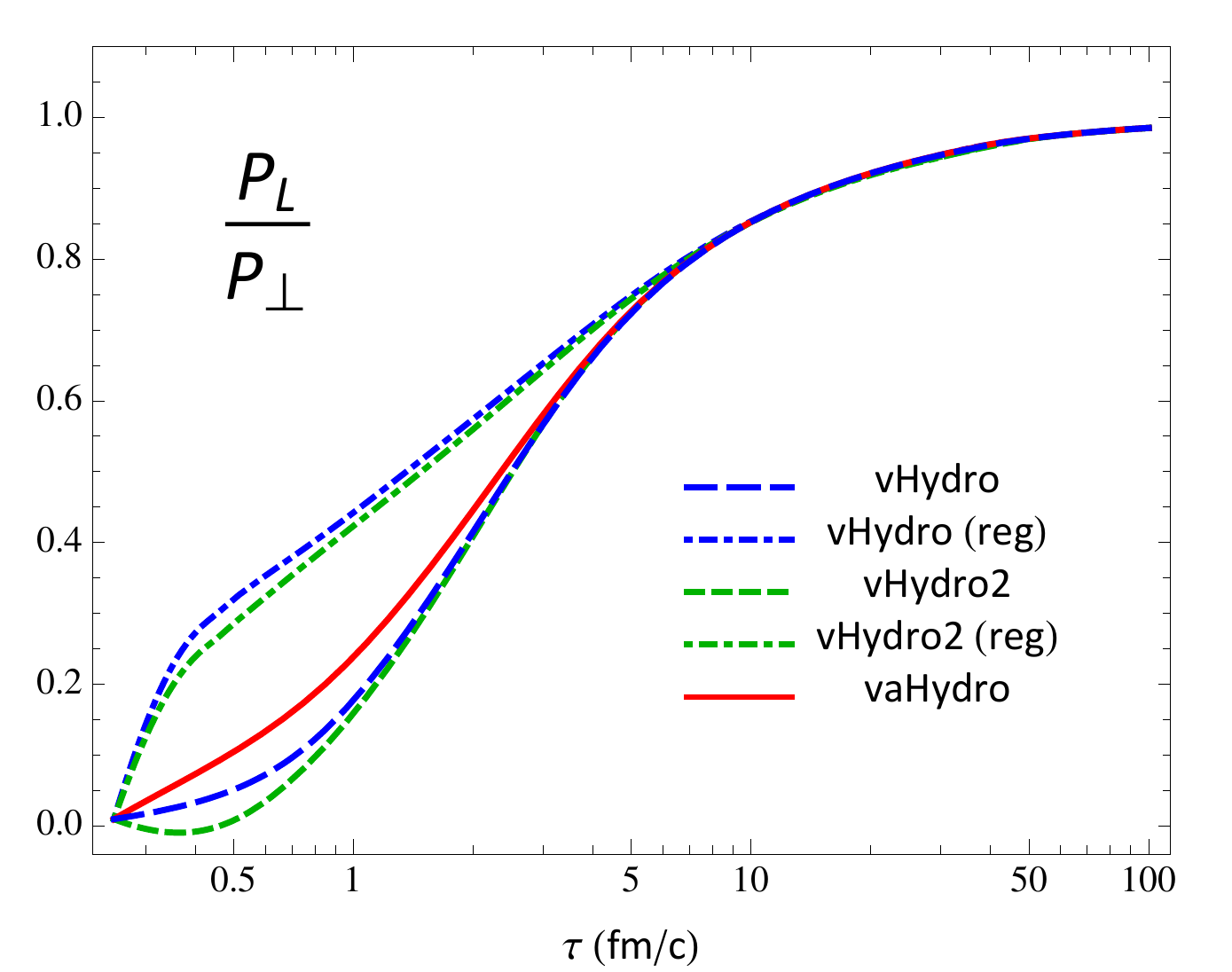}\hspace*{5mm}
\caption{
Bjorken evolution of the ratio $\PL / \Pperp$ for anisotropic hydrodynamics (solid red) and standard $2^{\text{nd}}$-order viscous hydrodynamics with quasiparticle (long-dashed blue) and small mass (long-dashed green) transport coefficients. Initial conditions are $T_0 = 0.5$\,GeV and $(\PL/\Pperp)_0 = 0.01$ at $\tau_0{\,=\,}0.25$\,fm/$c$. The regulated $\PL /\Pperp$ ratios (dash-dotted lines) are computed with the regulation scheme implemented in iEBE-VISHNU \cite{Shen:2014vra}, using a time step $\Delta\tau = 0.01$\,fm/$c$ in the hydrodynamic evolution. (We note that MUSIC \cite{Denicol:2018wdp} uses a less aggressive regulation scheme that effectively applies this type of regulation only to regions outside the freeze-out surface.)
\vspace*{7mm}
}
\label{fig3}
\end{SCfigure}

\vspace*{-7mm}
\section{Comparison to standard $2^{\text{nd}}$-order viscous hydrodynamics}
\label{sec4}

Figure 3 shows the Bjorken evolution of $\PL / \Pperp$ for anisotropic hydrodynamics (solid red) with Glasma-like initial conditions: $(\PL/\Pperp)_0 = 0.01$ and $T_0 = 0.5$\,GeV at $\tau_0 = 0.25$ fm/$c$. We compare anisotropic hydrodynamics to standard $2^{\text{nd}}$-order viscous hydrodynamics with quasiparticle (long-dashed blue) and small mass (long-dashed green) transport coefficients. For early times $\tau{\,\lesssim\,}1$\,fm/$c$, the $\PL / \Pperp$ ratio is slightly larger (corresponding to less anisotropy) in anisotropic than in standard viscous hydrodynamics.

In standard viscous hydrodynamics, the shear stress is very large at $\tau_0$, $\pi\simeq\Peq$. Standard viscous hydrodynamic codes usually require regulation of such large shear stresses for numerical stability, especially if they occur in the dilute periphery of the collision fireball. (Even there the (0+1)-d Bjorken model is a good approximation at very early times.) The need for such regulation is enhanced for small time resolution (e.g. $\Delta \tau = 0.01$\,fm/$c$), necessary when using fine transverse grids to resolve large initial spatial gradients. As shown by the dash-dotted lines in Fig.~\ref{fig3}, this regulation quickly (but unphysically) decreases the large initial pressure anisotropy. In anisotropic hydrodynamics, the pressure anisotropy is evolved non-perturbatively at leading-order, avoiding the appearance of large viscous corrections that may require regularization. The residual shear stresses in anisotropic hydrodynamics are much smaller than the longitudinal-transverse pressure difference $\PL{-}\Pperp$ and hopefully do not require significant regulation. 

\bibliographystyle{elsarticle-num}
\bibliography{QM18}

\end{document}